
\font\sixrm=cmr6           \font\sevenrm=cmr7
\font\tenrm=cmr10
\font\sixi=cmmi6           
\font\teni=cmmi10
\font\sixsy=cmsy6          
\font\tensy=cmsy10
   \font\sevenit=cmti7
\font\tenit=cmti10
   \font\sevensl=cmsl8 at 7pt
\font\tensl=cmsl10
          \font\sevenbf=cmbx7
\font\tenbf=cmbx10
   \font\seventt=cmtt8 at 7pt
\font\tentt=cmtt10

\font\twelverm=cmr12             \font\ninerm=cmr9
             \font\ninei=cmmi9
    \font\ninesy=cmsy9
            \font\nineit=cmti9
            \font\ninesl=cmsl10 at 9pt
            \font\ninebf=cmbx9
            \font\ninett=cmtt10 at 9pt

\font\fiverm=cmr5               \font\fourrm=cmr5 at 4pt
\font\fivei=cmmi5               \font\fouri=cmmi5 at 4pt
\font\fivesy=cmsy5              \font\foursy=cmsy5 at 4pt


\font\sevenex=cmex10 at 7pt
\def\tenpoint{%
\def\rm{\fam0\tenrm}%
\def\it{\fam\itfam\tenit}%
\def\sl{\fam\slfam\tensl}%
\def\bf{\fam\bffam\tenbf}%
\def\tt{\fam\ttfam\tentt}%
 \textfont0=\tenrm   \scriptfont0=\sixrm \scriptscriptfont0=\fiverm
 \textfont1=\teni    \scriptfont1=\sixi  \scriptscriptfont1=\fivei
 \textfont2=\tensy   \scriptfont2=\sixsy \scriptscriptfont2=\fivesy
 \textfont3=\tenex   \scriptfont3=\tenex \scriptscriptfont3=\tenex
 \textfont\itfam=\nineit
 \textfont\slfam=\ninesl
 \textfont\bffam=\tenbf
 \textfont\ttfam=\ninett
 \baselineskip=12pt
}
\def\ninepoint{%
\def\rm{\fam0\ninerm}%
\def\it{\fam\itfam\nineit}%
\def\sl{\fam\slfam\ninesl}%
\def\bf{\fam\bffam\ninebf}%
\def\tt{\fam\ttfam\ninett}%
 \textfont0=\ninerm   \scriptfont0=\fiverm \scriptscriptfont0=\fourrm
 \textfont1=\ninei    \scriptfont1=\fivei  \scriptscriptfont1=\fouri
 \textfont2=\ninesy   \scriptfont2=\fivesy \scriptscriptfont2=\foursy
 \textfont3=\sevenex  \scriptfont3=\sevenex
\scriptscriptfont3=\sevenex
 \textfont\itfam=\sevenit
 \textfont\slfam=\sevensl
 \textfont\bffam=\sevenbf
 \textfont\ttfam=\seventt
 \baselineskip=12pt
}
\font\tensmc=cmcsc10


\hsize     = 128mm
\vsize     = 190mm
\topskip   =  12pt
\leftskip  =  0mm
\parskip   =   0pt
\parindent =   4mm
\hoffset = 18mm
\voffset = 26mm

\def\Raggedright{%
 \leftskip=0pt
 \rightskip=0pt plus \hsize
 \spaceskip=.3333em
 \xspaceskip=.5em}

\def\Fullout{
 \leftskip=0pt
 \rightskip=0pt
 \spaceskip=0pt
 \xspaceskip=0pt}

\newcount\notenumber

\def\note{\global\advance\notenumber by 1
  \footnote{${^\the\notenumber}$}}

\newcount\authornumber

\def\authn{\global\advance\authornumber by 1\relax}

\newcount\titlerows

\def\titlen{\global\advance\titlerows by 1\relax}

\newcount\firstpageno
\newcount\lastpageno
\newcount\tabno

\newcount\figno

\def\rheadl{0}
\def\rheadr{0}
\def\studia{\sevenrm\frenchspacing Studia geoph. et geod. 42 (1998)}
\def\geos{\sevenrm\frenchspacing\copyright\ 1998 StudiaGeo s.r.o.,
Prague}
\def\blank{\quad\hfil\quad}
\def\revision{\bigskip
\line{\nineit Manuscript received: 31st January, 1998;\hfil
              Revisions accepted: 11th May, 1998}}


\def\ct#1\par{
 \twelverm\baselineskip=14pt
 \centerline{\uppercase{#1}}
 \titlen
 \ifnum\titlerows=1
 {\global\def\rhdr{\nineit #1}}{\global\def\rheadr{\nineit #1}}
 \else{\global\def\rheadr{\rhdr{\nineit\ ...}}}\fi}

\def\ca#1\par{
 \medskip
 \tensmc\baselineskip=12pt
 \centerline{#1}
 \authn
 \ifnum\authornumber=1
 {\global\def\rhdl{\nineit #1}}{\global\def\rheadl{\nineit #1}}
 \else
  \ifnum\authornumber=2{\global\def\rheadl{\rhdl{\nineit\ and\ #1}}}
  \else {\global\def\rheadl{\rhdl {\nineit\ et al.}}}\fi
 \fi}

\def\aa#1#2\par{
 \nineit\baselineskip=12pt
 \centerline{#1\note{\ninerm Address:\ #2}}
 }

\def\abstract#1\par{
\bigskip
 {\ninerm S u m m a r y :\ }
 \ninepoint\it\baselineskip=12pt
 #1\par}

\def\keywords#1\par{
\bigskip
\ninepoint\rm {K e y\quad w o r d s :\ }
 #1\par}

\def\ha#1\par{
 \goodbreak
 \Raggedright
 \tenrm\baselineskip=15pt
 \bigskip
 \centerline{\uppercase{#1}}
 \medskip}

\def\hb#1\par{
 \goodbreak
 \Raggedright
 \tenrm\baselineskip=15pt
 \bigskip
 \centerline{#1}
 \medskip}

\def\tx{
 \Fullout
 \baselineskip12pt
 \tenpoint\rm }

\def\bb#1\par{
  \hangindent=8mm
  \hangafter=1
  \frenchspacing
   \Fullout
 \ninepoint\rm
 \baselineskip=10pt
 \medskip
   \noindent{#1}\par}

\def\ref{\bigskip 
 \centerline{\nineit References}}

\def\tab#1\par{\bigskip 
{\global\advance\tabno by 1 \relax}
\ninepoint\rm
\baselineskip=10pt
\goodbreak
\noindent
{Table\ \number\tabno.\ #1}\smallskip}

\def\fig#1\par{
{\global\advance\figno by 1 \relax}
\ninepoint\rm
\baselineskip=10pt
\smallskip
\noindent
{Fig.\ \number\figno.\  #1}\bigskip}

\def\acknowledge#1\par{
\ninepoint\rm
\baselineskip=10pt
\medskip
{\it Acknowledgements:\ }#1\par}


\def\sethead{
\headline{\ifnum\pageno=\firstpageno\ \hfil\else
\ifodd\pageno\centerline{\rheadr}\else
\ifnum\pageno=\lastpageno\centerline{\rheadl :\ \rheadr}\else
\centerline{\rheadl}\fi
\fi\fi}}

\def\setfoot{
\footline{\ifodd\pageno\studia\ifnum
\firstpageno=\pageno,\ \number\firstpageno-\number\lastpageno\else
\relax\fi\hfil{\twelverm\folio}\else{\twelverm\folio}\hfil\studia\ifnu
\firstpageno=\pageno,\ \number\firstpageno-\number\lastpageno\else
\relax\fi\fi}}

\def\setfootm{
\advance\vsize by -1\baselineskip
\def\makefootline{\lineskip=12pt\baselineskip=10pt
 \vbox{\raggedright\noindent\ifodd\pageno\studia\ifnum
\firstpageno=\pageno,\ \number\firstpageno-\number\lastpageno\else
\relax\fi\hfil{\twelverm\folio}\else{\twelverm\folio}\hfil
\studia\ifnum
\firstpageno=\pageno,\ \number\firstpageno-\number\lastpageno\else
\relax\fi\fi\ifnum\firstpageno=\pageno\break\geos\hfil
\else\break\line{\blank}\fi
}}}

\firstpageno=1
\lastpageno=6
\nopagenumbers
\pageno=\firstpageno

\ct Dynamics of Axisymmetric Truncated Dynamo Models\par
\ca E. Covas and R. Tavakol\par
\aa {School of Mathem. Sci., Queen Mary and Westfield College,
London, UK}
{Mile End Road, London E1 4NS (eoc@maths.qmw.ac.uk)}\par

\sethead
\setfootm

\abstract  An  important  question regarding the study of mean field
dynamo models is how to make precise the nature of their  underlying
dynamics.  This is difficult both because relatively little is known
about  the  dynamical  behaviour of infinite dimensional systems and
also due to the numerical  cost  of  studying  the  related  partial
differential equations.

As  a  first  step  towards  their  understanding,  it  is useful to
consider the corresponding truncated models.  Here we summarise some
recent results of the study of a  class  of  truncated  axisymmetric
mean  field  dynamo  models.   We  find conclusive evidence in these
models for various  types  of  intermittency  as  well  as  multiple
attractors and final state sensitivity.

We  also  find  that the understanding of the underlying dynamics of
such dynamo models requires the study of a new  class  of  dynamical
systems,  referred to as the {\it non-normal} systems.  Current work
demonstrates that these types of systems are capable of a novel type
of intermittency and also of relevance for the understanding of  the
full axisymmetric PDE dynamo models.

\keywords  Axisymmetric mean field dynamos, intermittency, dynamical
systems

\ha 1. Introduction\par

\tx  There  is  proxy  evidence  suggesting variability in the solar
magnetic field over intermediate time  scales  of  order  of  $10^2$
years  ({\it Eddy, 1976}).  This type of variability is also thought
to be shared by solar type stars ({\it Baliunas et al., 1995}).  Now
given that there are no natural mechanisms with similar time  scales
operating  in  the  Sun and solar type stars ({\it Gough, 1990}) the
question arises as to the possible mechanisms responsible  for  such
variabilities.

There  is some observational evidence for the presence of non-linear
phenomena in stellar and solar magnetic activity.  As a  result  one
of  the  explanations  put forward over the last two decades is that
variability on these intermediate  time  scales  may  be  a  natural
outcome  of  the  nonlinear  regimes  operative  in such stars ({\it
Tavakol, 1978; Zeldovich et al., 1983; Weiss et al.,  1984;  Spiegel
et al., 1993}).

A  great  deal  of  effort  has  subsequently gone into the study of
magnetohydrodynamical dynamos operating  in  the  stellar  interiors
which  are  thought to give rise to such variability.  The equations
modelling these dynamo regimes are  nonlinear  partial  differential
equations (PDE).  A number of approaches have been employed to study
the  behaviour of such systems, including numerical studies of these
equations ({\it Brandenburg et al.,  1989a,  1989b;  Brandenburg  et
al.,  1990;  Brandenburg  et  al.,  1995}),  as well as their finite
dimensional truncations ({\it Zeldovich et al., 1983; Weiss et  al.,
1984;  Spiegel  et  al.,  1993}).   The difficulties with the former
models are twofold:  firstly the limits  imposed  by  the  numerical
cost  of their integration and secondly the fact that such numerical
results do not immediately make the dynamical mechanisms  underlying
them  transparent.   Such  understanding  is crucial if one hopes to
construct precise statistical measures in order to make  comparisons
with observations.

Now given that dynamical systems theory is well developed for finite
dimensional  flows, the latter (despite their approximate nature and
hence their limited direct physical applicability) are of  potential
value in making precise the underlying dynamics of such PDE models.

Here   we   summarise   some   recent   results({\it  Covas  et  al,
1997a,b,c,d}) which employ such truncations of the mean field dynamo
equations.

\ha 2. Model\par

\tx Our starting point is the mean  field  equation  describing  the
evolution of the mean magnetic field $\bf B$ in the form
$$
{\partial{\bf B}\over\partial t}\,=\,\nabla\times({\bf u}
\times{\bf B}\,+\,\alpha{\bf B}\,-\,\eta_t\nabla\times{\bf B}),
\eqno(1)
$$
where  ${\bf  u}$ is the mean velocity and $\eta_t$ is the turbulent
magnetic  diffusivity.   The  $\alpha$--effect   arises   from   the
correlation  of small scale turbulent velocities and magnetic fields
and is important in maintaining the dynamo action  by  relating  the
mean  electrical  current  arising in helical turbulence to the mean
magnetic field ({\it Krause and R\"adler, 1980}).

In  our studies we have considered the dynamical case where $\alpha$
can be divided into  a  hydrodynamic  ($\alpha_h$)  and  a  magnetic
part  ($\alpha_m$),  where the magnetic part satisfies an explicitly
time dependent diffusion type equation  with  a  nonlinear  forcing,
where  the  nonlinearity  is  quadratic in the magnetic field, given
by {\it Zeldovich et al.  (1983) {\rm and}  Kleeorin  and  Ruzmaikin
(1982)} (see {\it Covas et al., 1997a,b; 1998a} for details).

To  get  an understanding of the underlying dynamics of these models
we looked, using a spectral expansion, at the truncations  of  these
equations, which in a symbolic form can be written as
$$
{d{\bf x}\over dt}\,=\,\cal{F}({\bf x}, D, \nu)
\eqno(2)$$
where   ${\bf   x}  \in  \cal{R}^{T_d}$,  $T_d$  is  the  truncation
dimension, $D$ is the dynamo number,  $\nu$  is  the  ratio  of  the
turbulent  diffusivity  and  the  turbulent magnetic diffusivity and
$\cal{F}$ is a differentiable function which depends on  the  nature
of the equations and $T_d$.

A  great  deal  of  effort  has gone into the study of the truncated
dynamo models (see for example {Zeldovich et  al.,  1983;  Weiss  et
al.,  1984;  Spiegel  et  al.,  1993}).   Our aim has been to take a
systematic look at such truncated models firstly in  order  to  find
out   explicitly  what  types  of  complicated  nonlinear  modes  of
behaviour can occur in these models and secondly to find the general
features of the  appropriate  theoretical  framework  necessary  for
their  study  which we hope will in turn throw some light on the the
study of the full PDE dynamo models.

In the following section we summarise recent work from {\it Covas et
al.  (1997a,b; 1998a)}.  In particular we concentrate on  two  types
of  nonlinear  phenomena we have observed in such truncations, which
could in principle have potential relevance for  observed  behaviour
of solar type stars:  various types of intermittency and final state
sensitivity.

\ha 3. Results

\tx  Our  studies  of the above truncated models involved a range of
truncation  levels  $T_d$.   We  concentrated  on  two  families  of
truncated  models which had the least number of dimensions for which
a similar asymptotic behaviour  existed  (see  {\it  Covas  et  al.,
1997a}):\hfill\break
1.  Model I:  a six dimensional ($T_d=6$) truncated  model  of  type
(2)   which   is   antisymmetric   with   respect   to  the  equator
and\hfill\break
2.  Model II:  a twelve dimensional ($T_d=12$)  truncated  model  of
type  (2)  which  has  the  Model  I  as its antisymmetric invariant
subspace.\hfill\break
The following is a brief discussion of results.

\hb 3.1~~V a r i o u s~~~f o r m s~~~o f~~~i n t e r m i t t e n c y

\tx  {\bf  Crisis-induced Intermittency}.  Consider a system of type
(2) and let $D=D_c$ be  a  critical  dynamo  number  such  that  for
$D<D_c$  the  system possesses two chaotic attractors.  If as $D$ is
increased  the  attractors  enlarge  such  that  at   $D=D_c$   they
simultaneously  touch the boundary separating their basins, then for
$D$ slightly greater than $D_c$ a  typical  orbit  will  spend  long
periods  of  time  in  each  of  the  regions where the two previous
attractors existed and intermittently  switch  between  them.   This
form  of  dynamical  behaviour is referred to as {\it crisis-induced
intermittency}({\it Grebogi et al., 1987}).

We have found the presence of such a behaviour in the Model  I  with
$D_c=204.2796$ and $\nu=0.4$.  In these regimes, the distribution of
the  average time of reversals scales as a power law with $(D-D_c)$.
We have been able to substantiate this scaling in this  model  ({\it
Covas and Tavakol, 1997}).

{\bf  Type  I  Intermittency.} This type of behaviour is produced in
the neighbourhood of parameter values  where  a  periodic  orbit  is
destroyed  by  collision  with  an  unstable  orbit in a saddle node
bifurcation.  We have found the presence of such a behaviour in  the
Model  II  at  $D=170.0$  and  $\nu=0.5$.  A time plot of the energy
demonstrating this type of behaviour is depicted in Figure 1.

\midinsert
\vbox{\includegraphics{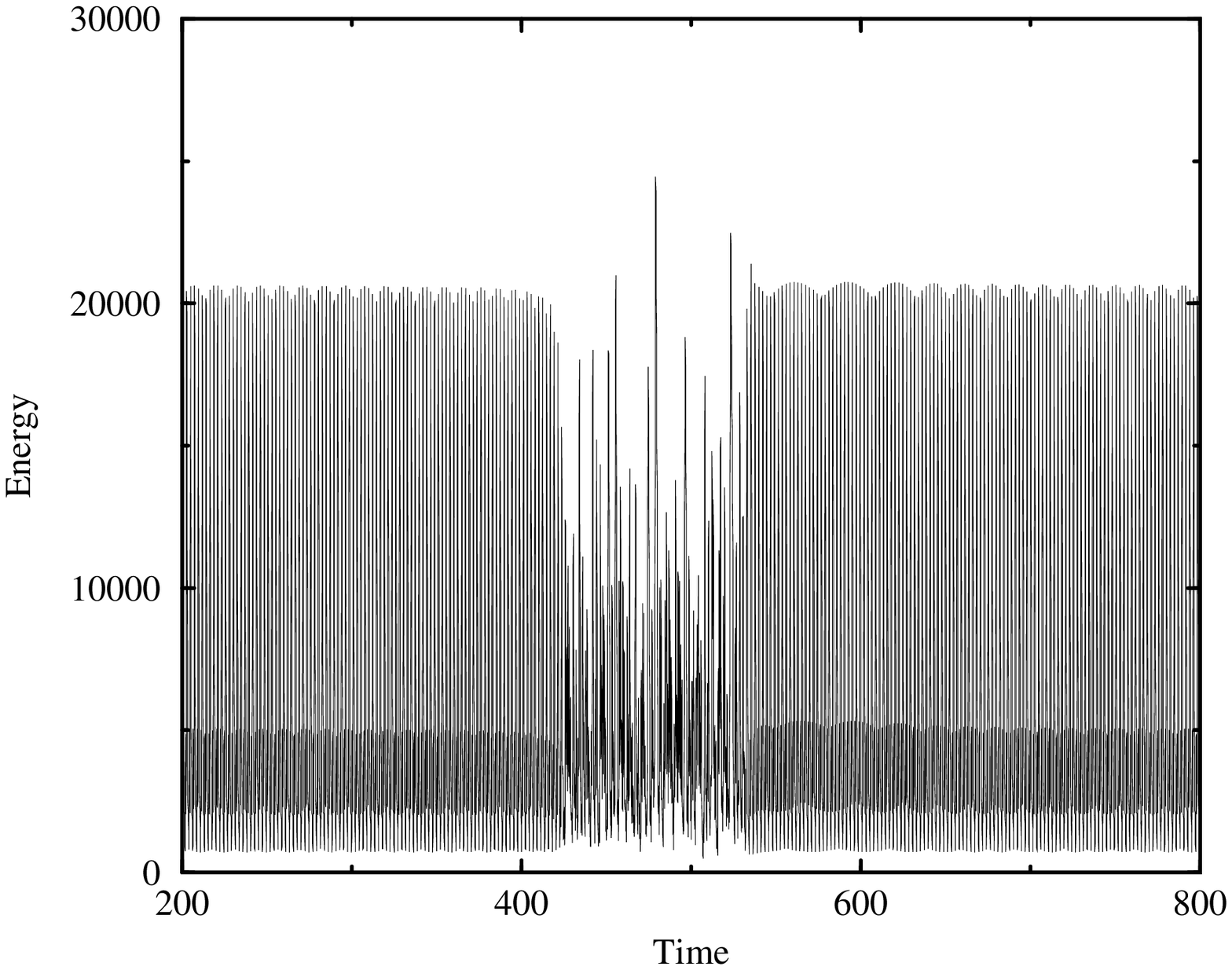}}
\vskip10cm
\fig Type I intermittency in Model II.  Note  the  decay  of  energy
when   the  burst  of  intermittency  occurs.   These  events  occur
repeatedly and in a irregular fashion.\par
\endinsert

{\bf On-Off Intermittency.} An important  feature  of  systems  with
symmetry  (as  in  the  case  of  solar  and stellar dynamos) is the
presence of invariant submanifolds.  It may happen  that  attractors
in  such  invariant  submanifolds  may become unstable in transverse
directions.  When this happens, one possible outcome could  be  that
the  trajectories can come arbitrarily close to this submanifold but
also have intermittent large  deviations  from  it.   This  form  of
intermittency  is  refereed as {\it on-off intermittency}({\it Platt
et al., 1993}).  Recent work has  shown  that  this  is  in  fact  a
generalised  form  of  on--off  intermittency  which  we refer to as
in--out intermittency ({\it Ashwin et al., 1998}).

We have found the presence of such a behaviour in the  Model  II  at
$D=177.70$ and $\nu=0.47$ ({\it Covas et al., 1997d}).

\hb 3.2~~F i n a l~~~s t a t e~~~s e n s i t i v i t y

\tx An important feature of  nonlinear  systems  is  that  they  can
possess {\it multiple attractors} at same parameter values.  In that
case,  the final outcome would depend upon the precise values of the
initial conditions.  It may turn out that the basins  of  attraction
could  have  a  fractal  or riddled nature, in which case very small
changes in the initial conditions can change  the  attractor,  hence
changing the final outcome of the dynamics.  In that case the system
is  said  to  have  {\it  final state sensitivity}({\it Alexander et
al.,1992; Lai and Grebogi 1996;  Lai  et  al.,  1996;  Ott  et  al.,
1994}).

We have found the presence of multiple attractors and fractal basins
for  the Model I at $D=204.2327$ and $\nu=0.5$.  This is depicted in
Figure 2.  Since  it  is  not  possible  to  determine  the  initial
conditions  precisely, this can lead to parameter regions where only
probabilistic statements may be made regarding the behaviour of such
stars.

\midinsert
\vbox{\includegraphics{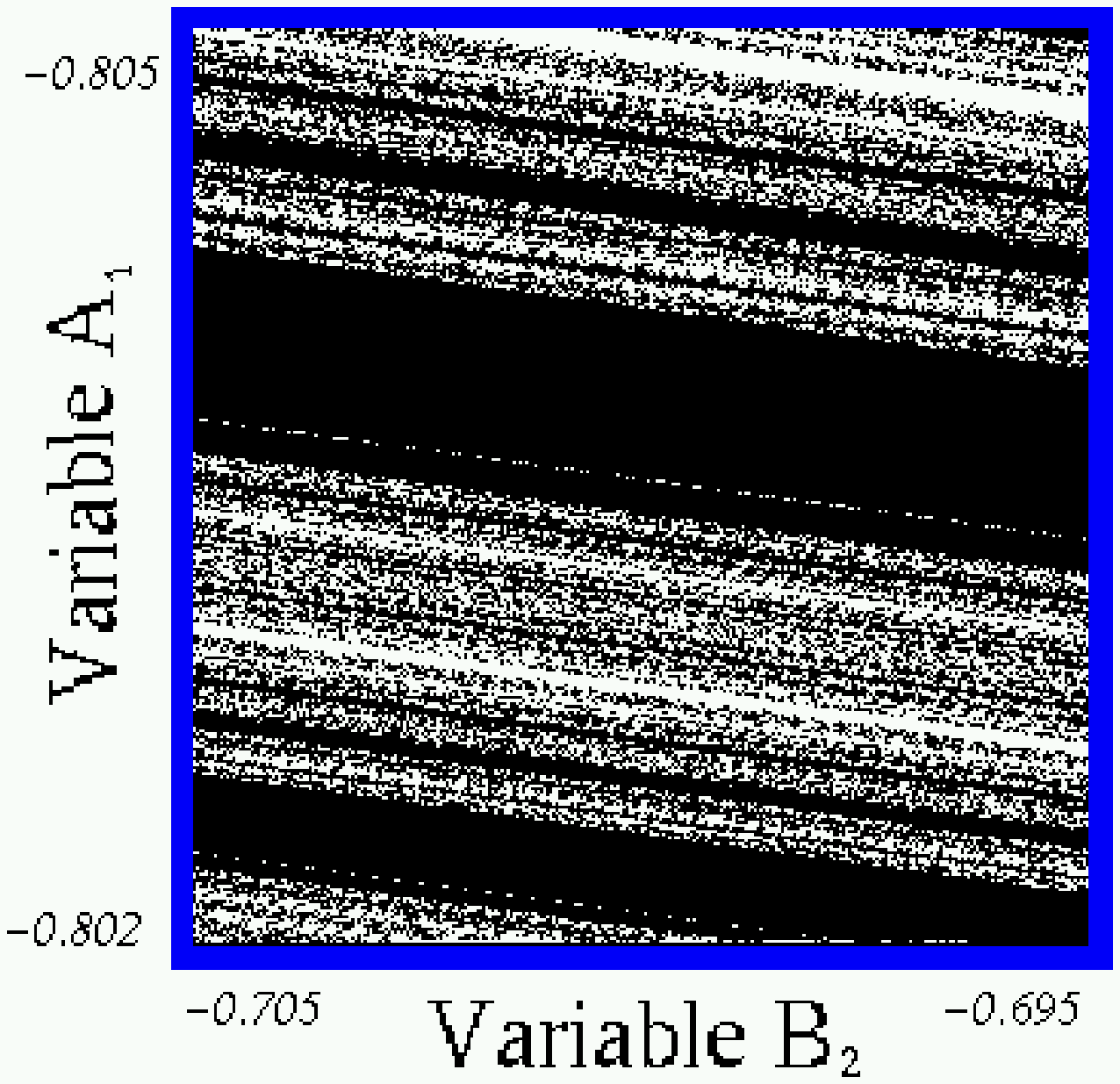}}
\vskip11cm
\fig Basin of attraction for Model I.  This is a slice over  two  of
the variables showing the basin of the two chaotic attractors at the
dynamo number $D=204.2327$.  This parameter value is just before the
crisis, where both attractors, and their corresponding basins, merge
with each other.\par
\endinsert

\ha 4. Discussion

\tx  As  a step towards understanding the underlying the dynamics of
solar and stellar variability and their corresponding PDE models, we
have  studied  the  detailed  dynamical  behaviour  using  truncated
axisymmetric  mean  field  dynamo  models.  We have found conclusive
evidence for the occurrence of a number of novel types of  behaviour
which  could  be of potential importance for the full dynamo models.
These are

1.  Various types of intermittency.  If repeated  in  real  dynamos,
these  types  of behaviour could be of significance in understanding
the Maunder type Minima.  We note that similar  types  of  behaviour
have  also been observed in numerical solutions of mean field dynamo
models ({\it Tworkowski et al., 1998; Covas et al., 1998a,b; Brooke,
1997; Brooke et al., 1998}).

2.  Final state sensitivity.  The presence of  such  sensitivity  in
real  stars  could be of significance in that, for example, it could
potentially lead to stars of same spectral type, rotational  period,
age   and   compositions   showing   different  modes  of  dynamical
behaviour.  There is some evidence for similar types of behaviour in
PDE  models  ({\it   Tavakol   et   al.,   1995;   Covas   et   al.,
1998a}).

3.   Our studies show that an important feature of dynamo models and
their truncations is that their  parameters  are  {\it  non-normal},
i.e.   they  vary the dynamics within the invariant subspace as well
as outside it.  In fact we expect such non-normality to be a generic
property of parameters in general truncations of  physical  systems.
A  framework  for  understanding  the  dynamics  of  such systems is
developed elsewhere ({\it Covas et al., 1997; Ashwin et al.,  1998})
and a detailed study of the relationship between these phenomena and
the PDE models is to appear elsewhere ({\it Covas et al., 1998b}).

\acknowledge We wish to thank Axel Brandenburg and Andrew Tworkowski
for  many  useful  conversations.   EC wishes to thank the Astronomy
Unit at  QMW  and  the  organisers  of  the  5$^{th}$  International
Workshop on {\it Planetary and Cosmic Dynamos} for support to attend
the  conference  held  in  Trest  1997.   EC  is  supported by grant
BD/5708/95 -- Program  PRAXIS  XXI,  from  JNICT  --  Portugal.   RT
benefited  from  PPARC  UK  Grant  No.   L39094.  This research also
benefited from the EC Human Capital and  Mobility  (Networks)  grant
``Late   type   stars:    activity,   magnetism,   turbulence''  No.
ERBCHRXCT940483.

\revision
\ref

\bb Alexander, J., Kan, I., Yorke, J.  and  You.,  Z.,  1992: Riddled Basins.  {\it
Int.  Journal of Bifurcations and Chaos} {\bf 2} 795.

\bb   Ashwin,   P.,   Covas,   E.   and  Tavakol,  R.,  1998:  Transverse instability for non-normal parameters. {\it
Nonlinearity}, submitted

\bb Baliunas, S.L.  et al., 1995: Chromospheric variations in main--sequence starts. II.  {\it Ap.J.} {\bf 438}, 269.

\bb Brandenburg, A., Krause, F., Meinel, R., Moss, D.  and Tuominen,
I., 1989: The stability of nonlinear dynamos and the limited role of kinematic growth 
rates. {\it Astron.  and Astrophys} {\bf 213}, 411.

\bb Brandenburg, A., Moss, D.  and Tuominen, I., 1989: Nonlinear dynamos with magnetic buoyancy in spherical geometry. {\it  Geophys.
Astrophys.  Fluid Dyn.} {\bf 40}, 129.

\bb  Brandenburg,  A.,  Moss,  D.,  R\"udiger, G.  and Tuominen, I.,
1990: The nonlinear solar dynamo and differential rotation: A Taylor
number puzzle? {\it Solar Physics} {\bf 128}, 243.

\bb  Brandenburg,  A.,  Nordlund,  A.,  Stein, R.F.  and Torkelsson,
1995: Dynamo-generated turbulence and large-scale magnetic fields in a Keplerian
shear flow. {\it ApJ} {\bf 446}, 741.

\bb Brooke, J.  M., 1997: Breaking of equatorial symmetry in a rotating system: a spiralling 
intermittency mechanism. {\it Europhysics Letters} {\bf 37}, 171.

\bb Brooke, J.M., Pelt, J., Tavakol, R.\ and Tworkowski, A., 1998:
Grand minima and equatorial symmetry breaking in axisymmetric dynamo
models. {\it Astronom. and Astrophys.} {\bf 332}, 339.

\bb  Covas,  E.,  Tworkowski,  A., Brandenburg, A.  and Tavakol, R.,
1997a: Dynamos with different formulations of a dynamic $\alpha$--effect. {\it Astronom.  and Astrophys.} {\bf 317}, 610.

\bb Covas, E., Tworkowski, A., Tavakol,  R.   and  Brandenburg,  A.,
1997b: Robustness of truncated $\alpha \omega$ dynamos with a         
dynamic $\alpha$. {\it Solar Physics} {\bf 172}, 3.

\bb  Covas,  E.  and Tavakol, R., 1997c: Crisis in truncated mean field dynamos. {\it Phys.  Rev.  E.} {\bf
55}, 6641.

\bb Covas, E., Ashwin, P.   and  Tavakol,  R.,  1997d: Non-normal parameter blowout bifurcation in a truncated mean field dynamo.  {\it  Phys.
Rev.  E.} {\bf 56}, 6451.

\bb  Covas,  E.,  Tavakol,  R.,  Tworkowski, A.\ \& Brandenburg, A.,
1998a: Axisymmetric mean field dynamos with dynamic and algebraic
$\alpha$--quenchings. {\it Astron.  and Astrophys.} {\bf 329}, 350.

\bb Covas, E., Tavakol, R., Ashwin, P., Tworkowski, A.  and  Brooke,
J.M., 1998b: In--out intermittency in PDE and ODE models of axisymmetric mean-field
dynamos. {\it Phys. Rev. Lett.}, submitted.

\bb Eddy, J. A., 1976: The Maunder Minimum {\it Science} {\bf 192}, 1189.

\bb  Gough, D.  O., 1990: On possible origins of relatively short--term variations
in the solar structure. {\it Phil.  Trans.  R.  Soc.  Lond.} {\bf
A330}, 627.

\bb Grebogi, C., Ott, E., Romeiras, F., and Yorke, J.A., 1987: Critical exponents for crisis--induced intermittency. {\it
Phys.  Rev A.}, {\bf 36}, 5365.

\bb   Kleeorin,   N.    I.    and   Ruzmaikin,   A.A.,  1982: Dynamics of
the average turbulent helicity in a magnetic field.  {\it
Magnetohydrodynamica} {\bf N2}, 17.

\bb Kleeorin, N.  I., Rogachevskii, I.   and  Ruzmaikin,  A.,  1995: Magnitude of the dynamo--generated magnetic field in solar--type
convective zones.
{\it Astron.  and Astrophys.} {\bf 297}, 159.

\bb   Krause,  F.   and  R\"adler,  K.-H.,  1980:   {\it  Mean-Field
Magnetohydrodynamics and Dynamo Theory}, Pergamon, Oxford

\bb Lai, Y-C.  and Grebogi, C., 1996: Characterising riddled fractal sets. {\it Phys.   Rev.   E.}  {\bf
53}, 1371 (reference 3).

\bb  Lai,  Y-C.,  Grebogi, C., Yorke, J.A.  and Venkataramani, S.C.,
1996: Riddling bifurcation in chaotic dynamical systems. {\it Phys.  Rev.  Lett.} {\bf 77}, 55.

\bb Ott, E., Sommerer, J.C., Alexander, J., Kan, I.  and  Yorke,  J.
A., 1994: A transition to chaotic attractors with riddled basins. {\it Physica D} {\bf 76}, 384.

\bb  Platt,  N.,  Spiegel,  E.   and  Tresser, C., 1993: On-off intermittency; a mechanism for bursting. {\it Phys.
Rev.  Lett.} {\bf 70}, 279.

\bb Spiegel, E., Platt, N.  and Tresser, C.,  1993: The Intermittent Solar Cycle.  {\it  Geophys.
and Astrophys.  Fluid Dyn.} {\bf 73}, 146.

\bb  Tavakol, R.K., Tworkowski, A.S., Brandenburg, A., Moss, D.  and
Tuominen, I., 1995: Structural stability of axisymmetric
dynamo models. {\it Astron.  and Astrophys.} {\bf 296}, 269.

\bb Tworkowski, A., Tavakol, R.,  Brandenberg,  A.,  Moss,  D.   and
Tuominnen, I., 1998: Intermittent behaviour in
axisymmetric mean-field dynamo models in spherical shells. {\it Mon.  Not.  R.  Astr.  Soc.}, in press.

\bb  Weiss,  N.O.,  Cattaneo,  F.,  and  Jones,  C.A.,  1984: Periodic and aperiodic dynamo waves.  {\it
Geophys.  Astrophys.  Fluid Dyn.} {\bf 30}, 305.

\bb Zeldovich, Ya.B., Ruzmaikin, A.A.   and  Sokoloff,  D.D.,  1983:
{\it Magnetic Fields in Astrophysics}, Gordon and Breach, New York

\bye